\documentclass[12pt,preprint]{aastex} 

\usepackage{graphicx}
\usepackage[twocolumn]{emulateapj5}

\newcommand{\halpha}{H$\alpha$}
\newcommand{\hbeta}{H$\beta$}
\newcommand{\hgamma}{H$\gamma$}

\newcommand{\sameauthor}{\underbar{\qquad\qquad}.}
\newcommand{\usp}{\underbar{\hskip 7pt}}

\shortauthors{SCHMIDT ET AL.}
\shorttitle{SDSS MAGNETIC WHITE DWARFS}

\slugcomment{To appear in the Oct. 1, 2003 Astrophysical Journal}
\received{}
\revised{}

\begin{document}

\title{Magnetic White Dwarfs from the SDSS. The First Data Release\altaffilmark{1}}

\altaffiltext{1}{A portion of the results presented here were obtained with
the MMT Observatory, a facility operated jointly by The University of Arizona
and the Smithsonian Institution.}

\author{
Gary D. Schmidt\altaffilmark{2},
Hugh C. Harris\altaffilmark{3},
James Liebert\altaffilmark{2},
Daniel J. Eisenstein\altaffilmark{2},
Scott F. Anderson\altaffilmark{4},
J. Brinkmann\altaffilmark{5},
Patrick B. Hall\altaffilmark{6,7},
Michael Harvanek\altaffilmark{5},
Suzanne Hawley\altaffilmark{4},
S. J. Kleinman\altaffilmark{5},
Gillian R. Knapp\altaffilmark{6},
Jurek Krzesinski\altaffilmark{5,8},
Don Q. Lamb\altaffilmark{9},
Dan Long\altaffilmark{5},
Jeffrey A. Munn\altaffilmark{3},
Eric H. Neilsen\altaffilmark{10},
Peter R. Newman\altaffilmark{5},
Atsuko Nitta\altaffilmark{5},
David J. Schlegel\altaffilmark{6},
Donald P. Schneider\altaffilmark{11},
Nicole M. Silvestri\altaffilmark{4},
J. Allyn Smith\altaffilmark{12},
Stephanie A. Snedden\altaffilmark{5},
Paula Szkody\altaffilmark{4},
and
Dan Vanden Berk\altaffilmark{13}
}

\altaffiltext{2}{Steward Observatory, The University of Arizona, Tucson AZ
85721.} \email{gschmidt@as.arizona.edu}
\altaffiltext{3}{U.S. Naval Observatory, P.O. Box 1149, Flagstaff, AZ
86002-1149.}
\altaffiltext{4}{Department of Astronomy, University of Washington, Box 351580,
Seattle, WA 98195-1580.}
\altaffiltext{5}{Apache Point Observatory, PO Box 59, Sunspot, NM 88349-0059.}
\altaffiltext{6}{Princeton University Observatory, Peyton Hall, Princeton, NJ
08544-1001.}
\altaffiltext{7}{Departamento de Astronom\'{\i}a y Astrof\'{\i}sica, Facultad
de F\'{\i}sica, Pontificia Universidad Cat\'{o}lica de Chile, Casilla 306,
Santiago 22, Chile.}
\altaffiltext{8}{Obserwatorium Astronomiczne na Suhorze, Akademia Pedagogicazna
w Krakowie, ulica Podchor\c{a}\.{z}ych 2, PL-30-084 Krak\'{o}w, Poland.}
\altaffiltext{9}{Dept. of Astronomy and Astrophysics, University of Chicago,
5640 South Ellis Avenue, Chicago, IL 60637.}
\altaffiltext{10}{Fermi National Accelerator Laboratory, P.O. Box 500, Batavia,
IL 60510.}
\altaffiltext{11}{Pennsylvania State University, Department of Physics and
Astronomy, 525 Davey Lab., University Park, PA 16802.}
\altaffiltext{12}{Mailstop D448, Los Alamos National Lab, Los Alamos, NM 87545.}
\altaffiltext{13}{Department of Physics and Astronomy, University of
Pittsburgh, 3941 O'Hara Street, Pittsburgh, PA 15260.}
\begin{abstract}

Beyond its goals related to the extragalactic universe, the Sloan Digital Sky
Survey (SDSS) is an effective tool for identifying stellar objects with unusual
spectral energy distributions.  Here we report on the 53 new magnetic white
dwarfs discovered during the first two years of the survey, including 38 whose
data are made public in the 1500 square-degree First Data Release.  Discoveries
span the magnitude range $16.3 \le g \le 20.5$, and based on the recovery rate
for previously-known magnetic white dwarfs, the completeness of the SDSS
appears to be high for reasonably hot stars with $B \gtrsim 3$~MG and
$g\gtrsim15$.  The new objects nearly double the total number of known
magnetic white dwarfs, and include examples with polar field strengths
$B_p>500$~MG as well as several with exotic atmospheric compositions.  The
improved sample statistics and uniformity indicate that the distribution of
magnetic white dwarfs has a broad peak in the range $\sim$$5-30$~MG and a tail
extending to nearly 10$^9$~G.  Degenerates with polar fields $B_p\gtrsim50$~MG
are consistent with being descendents of magnetic Ap/Bp main-sequence stars,
but low- and moderate-field magnetic white dwarfs appear to imply another
origin.  Yet-undetected magnetic F-type stars with convective envelopes that
destroy the ordered underlying field are attractive candidates.

\end{abstract}


\section{Introduction}

Among the many products of the Sloan Digital Sky Survey (SDSS; York et al.
2000; Stoughton et al. 2002) will be thousands of new white dwarfs extending to
fainter than 20th mag. and distances greater than 1 kpc. Local samples show
that $\sim$10\% of white dwarfs are magnetic in the range $10^4-10^9$~G (Kawka
et al. 2003; Liebert, Bergeron \& Holberg 2003; see also Schmidt \& Smith
1995).  Thus, when the SDSS is complete to at least 7,000 square degrees, the
list of 60-odd previously-known magnetic white dwarfs (e.g., Wickramasinghe \&
Ferrario 2000) will be increased several times over.  The new catalog will
enable a number of statistical tests of roles that magnetic fields might play
in stellar evolution, and will almost certainly display new atomic and
molecular species in magnetic fields beyond what have been observed to date.  A
hint of what can be expected is provided by the 7 new magnetic white dwarfs
identified by G\"ansicke, Euchner, \& Jordan (2002) in the first 462
square-degree Early Data Release (EDR). The current paper reports on the 60
magnetic white dwarfs identified thus far in the SDSS; 53 of which are new and
38 of which are contained in the First Data Release (DR1; Adelman et al. 2003)
that covers the initial $\sim$1500 square degrees of the survey. Some of these
stars have been reported in the preview by Harris et al. (2003).

\section{Observations}

The primary SDSS database is 5-color photometry ($u,g,r,i,z$) from which
targets are selected for followup fiber spectroscopy with twin dual-beam
spectrographs ($3900-6200$~\AA\ and $5800-9200$~\AA).  Both sets of data are
obtained with the 2.5 m telescope at Apache Pt., New Mexico (see, e.g.,
Fukugita et al. 1996; Gunn et al. 1998; Lupton, Gunn, \& Szalay 1999; York et
al. 2000; Lupton et al. 2001).  Astrometric data to an accuracy of $\sim$50 mas
is also available (Pier et al. (2003), and provides important information for
local stellar populations like white dwarfs.  Target selection for the
spectroscopic fibers is a somewhat involved process based on color criteria
that have evolved slightly since the survey's inception.  A brief summary can
be found in Harris et al. (2003); more complete discussion will be deferred to
a future publication.  However, it is important to realize that only the
coolest white dwarfs are targeted explicitly.  Once the fiber data are reduced
to flux-calibrated spectra, the objects are crudely classified.  White dwarfs
can be found in a number of categories, but because of their occasionally
bizarre spectral characteristics, a search for magnetic white dwarfs must
involve a variety of object categories, including QSOs and especially
unclassifiable objects. Still, undoubtedly many white dwarfs do not receive a
fiber, and the resulting incompleteness of the spectroscopic sample cannot be
accurately assessed at this time.  With a resolving power of $\sim$1800, SDSS
spectroscopy is effective for recognizing magnetic splitting of the Balmer
lines for $B>1-3$~MG, depending on object brightness ($\Delta\lambda_Z \approx
\pm 20$~\AA\ MG$^{-1}$ at \halpha).  The detection of weaker fields on white
dwarfs requires high-resolution spectroscopy of the non-LTE line cores or a
search for circular polarization in the line wings arising from the net
longitudinal field component because the individual Zeeman lines are lost in
the Stark-broadened profiles.  Very strong magnetic fields can generally be
recognized by the presence of broad, generally asymmetric absorption features
at unusual wavelengths.

We also present here the initial results of followup optical spectroscopy and
circular spectropolarimetry being carried out for the unusual, more
interesting, and uncertain SDSS magnetic white dwarfs.  For the latter
category, a detection of circular polarization in the continuum serves to
confirm the presence of a strong magnetic field, since it is produced by
magnetic dichroism in the stellar atmosphere at a rate $\sim$0.01\% MG$^{-1}$
(Schmidt 1989), dependent also on temperature and field orientation. The
spectropolarimetric data were obtained during the period 2002 Feb. $-$ Dec.
with the instrument SPOL (Schmidt, Stockman, \& Smith 1992) attached to the
Steward Observatory 2.3 m Bok telescope on Kitt Peak and the 6.5 m MMT atop Mt.
Hopkins.  Data reduction was carried out as described by Schmidt et al.
(1992).

\section{Results}

Magnetic white dwarfs selected to date from the SDSS are summarized in Tables
$1-3$ for various spectral types.  In each table, column (1) lists the SDSS
identification in epoch 2000 coordinates, col. (2) the plate-MJD-fiber
identifier for the SDSS spectrum, and cols. (3) and (4) the UT date and time of
the spectroscopy.  Spectroscopic exposure times are typically $\sim$45 min in
length and are selected to achieve approximately uniform signal-to-noise ratio
for a specified object brightness.  Occasionally this requires multiple
exposures over successive nights, so rotational smearing of the field patterns
may be present, depending on the (unknown) spin period and details of the
observation.

Polar field strengths and approximate viewing perspectives have been estimated
where possible by comparing the SDSS data against model spectra for assumed
inclined dipolar field patterns, and these results are listed in cols. (6) and
(7). The modeling procedure is an outgrowth of the code developed by Latter,
Schmidt, \& Green (1987), using transition wavelengths and oscillator strengths
from Kemic (1974) for $B_p\lesssim30$~MG and the calculations summarized by
Ruder et al. (1994) for higher fields. The procedure accounts for the change in
line strength with field as well as the variation in $B$ over the stellar
surface, but does not solve for the temperature structure nor account for the
transfer of polarized radiation through the atmosphere.  This ``geometric''
approach is therefore adequate for estimating the effective polar field
strength and orientation for a simple assumed field pattern like a centered or
offset dipole, but not for evaluating the appropriateness of that geometry vs.
more elaborate multipolar expansions from detailed spectropolarimetric data.
An estimate of the reliability of the approach as compared with more elaborate
procedures can be gained by a comparison of our results against those of
G\"ansicke et al. (2002) for the stars in the EDR. Overall, we regard the
results quoted below to be accurate to better than 10\% in $B_p$ and
$\sim$$\pm$30$^\circ$ in inclination for the field patterns assumed. That said,
we note that the profiles of geometry-sensitive lines like the $\sigma$
components of \halpha\ are often not well modeled by a dipolar field pattern at
{\it any\/} assumed inclination; we take this to indicate that the field
patterns cannot be adequately described by simple centered dipoles. The final
column of the tables provides aliases for previously-known objects and other
comments, including the estimates of $B_p$ and $i$ by G\"ansicke et al. for
stars in common.

About 75\% of all white dwarfs are type DA, and the 49 stars with hydrogen
features in Table 1 also dominate the 60 magnetic candidates identified thus
far in the SDSS.  Field strengths of the stars span the range 1.5~MG to more
than 500~MG, nearly as wide an interval as is covered by all
previously-discovered magnetic white dwarfs. Caution must be exercised when
evaluating the number distribution as a function of field strength, however,
because of the aforementioned incompleteness at weak fields, and because the
existence of a correlation between magnetism and remnant mass (e.g., Liebert
1988; Sion et al. 1988) implies that selection effects may be important in a
magnitude-limited survey like the SDSS.  Such biases should be carefully
assessed when estimating relative space densities (Liebert et al. 2002).  We
return to this question in $\S$6.2.

A selection of white dwarfs from Table 1 showing the hydrogen spectrum in low
to moderate fields is shown in Figure 1.  In the regime $0<B\lesssim50$~MG,
\halpha\ retains a triplet structure with ever-increasing separation but with
the triplet components broadening as the $l$-degeneracy of the atom is
removed.  For considerably stronger fields, the magnetic effect can no longer
be considered a perturbation and diagnosis is best done by searching for
features at the locations of turnarounds or stationary points in the
$B-\lambda$ curves.  Examples of such stars with field strengths between 100~MG
and 500~MG are compared to the principal features expected in Figure 2.

Only 4 magnetic DB stars were known prior to the SDSS: LB8827 has
$B\lesssim1$~MG and HE1211$-$1707, HE1043$-$0502, and GD 229 are all above
50~MG. Additionally, G227-28 and Feige 7 are magnetic stars with mixed H/He
composition.  Three new magnetic DB white dwarfs with $B<10$~MG have been found
in the SDSS and are displayed in Figure 3. Such stars are most easily
recognized by the Zeeman triplet at \ion{He}{1} $\lambda$5876, which is
isolated and not easily confused with other features. These spectra have also
been modeled for dipolar field strength and inclination using synthetic spectra
based on the calculations of Kemic (1974), with the results listed in Table 2.

Two magnetic/nonmagnetic white dwarf pairs have been found in the SDSS that
augment the 3 binaries known previously.  Both magnetic stars are DAH spectral
type and entered in Table 1, but as depicted in Figure 4, they are teamed with a
nonmagnetic DA and DB.  Such systems are potentially very interesting from
evolutionary points of view, and the new examples are described individually
below.

Finally, Table 3 collects stars that show absorption lines or bands of metals
in magnetic fields, magnetic stars whose atmospheric species are as yet
unknown, and objects that can only be classified as candidate magnetic white
dwarfs at this time. These include the first DQ showing Zeeman-split \ion{C}{1}
lines and a magnetic DZ similar to LHS 2534 (Reid, Liebert, \& Schmidt 2001).
The spectra are displayed in Figures $5-6$ and discussed in more detail below.
Note that field strengths for these stars, where available, are quoted as mean
surface fields, $B_s$, since atomic data are generally insufficient to allow
detailed modeling.

\section{Photometry and Stellar Temperatures}

The SDSS spectroscopy is known to occasionally display flux calibration
irregularities due to variable fiber placement, atmospheric dispersion, and
other effects.  When present, these tend to be most prominent shortward of
4500~\AA, so the observed spectral flux distribution is often not ideal for
temperature estimation.  A more accurate approach -- the fitting of Balmer line
profiles for simultaneous temperature and gravity determination -- has been
applied only rarely to magnetic white dwarfs (e.g., Schmidt et al. 1992) due to
the uncertainties of line broadening, and a general approach has not yet been
developed.

We have therefore gathered the ``PSF photometry'' from the imaging
observations, using the beta version of the First Data Release (Adelman et al.
2003), specifically from version v5\_3 of the {\tt photo} pipeline.  Note that
this can be slightly different from the photometry upon which spectroscopic
target decisions were made.  The PSF photometry has systematic effects that are
$\lesssim$2\% for $g$ through $z$ and only slightly poorer in $u$ (Hogg et al.
2001; Smith et al., 2002; Stoughton et al. 2002).  The values are presented in
Table 4.

Temperatures have been estimated for the magnetic stars by comparing their
locations in the $u-g$ {\it vs.\/} $g-r$ plane against DA and DB colors
computed for nonmagnetic stars from the models of Bergeron, Wesemael, \&
Beauchamp (1995; see also Harris et al. 2003). These results are also provided
in Table 4.  A few possible systematic effects should be noted.  First, the
equivalent widths of magnetically-split absorption lines generally increase
with increasing field strength, and the patterns for the Balmer series begin to
overlap for $\lambda\lesssim4200$~\AA\ at $B\sim10$~MG.  This can result in a
depressed $u$ magnitude and an erroneously cool temperature. Second, the
magnitudes quoted are observed SDSS values; transformation coefficients for
placing the photometry onto the AB system are not yet finalized and no
reddening corrections have been made.  Both corrections would be in the
direction of making the colors bluer, with an effect on stellar temperatures of
as much as 10\%.  A larger systematic difference exists between the
temperatures quoted here compared with the continuum-based values of G\"ansicke
et al. (2002) for the same EDR magnetic white dwarfs; the latter are almost
certainly overestimates since the pipeline processing of SDSS spectroscopy
includes a reddening correction for an entire intervening Galactic disk.
Finally, several entries in Table 4 are annotated by a colon (:) to denote
unusually large uncertainties. These indicate the few cases where the stellar
colors fall outside the loci for typical white dwarf gravities, as well as
those examples with fields $\gtrsim100$~MG, where continuum opacities depart
substantially from the zero-field forms (and are unknown).  Of course,
temperatures for the two composite-spectrum objects represent the overall flux
distribution, and DA atmospheres were assumed in both cases.

\section{Notes on Individual Objects}

{\it SDSS J0042+0019 and SDSS J0847+4842:\/} 
Both of these stars are almost certainly magnetic+nonmagnetic
double-degenerate binaries. SDSS J0042+0019\footnote{For brevity, we refer to
Sloan targets as SDSS J$hhmm{\pm}ddmm$ in the text of this paper.} (Figure 4)
displays Zeeman-split Balmer lines suggesting a field of $B_p\approx14$~MG,
but the $\pi$ components are too strong compared with the $\sigma$ lines for
any reasonable field pattern.  We interpret the spectrum as a composite DA+DAH
with little radial velocity difference between the stellar components at the
time of observation.  SDSS J0847+4842 exhibits very broad lines of hydrogen plus
narrower, undisplaced features of \ion{He}{1}. Our followup
spectropolarimetry, shown in the same figure, supports the DAH+DB
classification, with strong $\cal S$-wave circular polarization signatures
across \hbeta\ and \hgamma\ but no polarization reversals in the helium
lines.  Without resolving the Balmer line cores, it is difficult to estimate
the surface field strength on the magnetic star; the magnitude of the
longitudinal Zeeman effect in polarization suggests a field of a few MG. This
inability to resolve the Balmer lines cores may also indicate a high rotation
velocity on the magnetic component, such as would result from accretion.
Further observations are warranted to determine the periods of these binaries.

{\it SDSS J0157+0033:\/} 
This is only the second magnetic DZ star known, showing lines of \ion{Mg}{1},
\ion{Na}{1}, and \ion{Ca}{1}.  As discussed by Reid et al. (2001) for LHS 2534
(which is recovered here as SDSS J1214$-$0234 and compared with SDSS J0157+0033
in Figure 5), the relative importance of magnetic and spin-orbit effects varies
among these features for a field strength of a few MG.  The \ion{Na}{1} D
lines, with a fine-structure splitting of 6~\AA, decouple first, near a field
of 1~MG.  Thus, a deep triplet centered at $\lambda$5893 in LHS 2534 was
interpreted in terms of the linear Zeeman effect and provided an estimate for
the mean field of $B_s=1.92$~MG. At that field strength, the magnetic and
spin-orbit effects are comparable for the \ion{Mg}{1} b lines, and four rather
evenly-spaced components were observed, displaced somewhat to the red of the
zero-field triplet.

Like LHS~2534, the spectrum of SDSS J0157+0033 shows a feature at
5893~\AA\ which we take to be the $\pi$ component of \ion{Na}{1} D in an
ordinary magnetic triplet.  A second line at 5834~\AA\ can then be interpreted
as the $\sigma^-$ component, implying $B_s=3.7$~MG. The predicted location of
the $\sigma^+$ component is indicated in Figure 5, but noise in the spectrum
prevents a clear identification.  The stronger field on SDSS J0157+0033 as
compared with LHS~2534 suggests that the \ion{Mg}{1} lines may have also
entered the Paschen-Back regime, where a linear interpretation is adequate.
Indeed, a simple model for $B=3.7$~MG reveals that the nine components have
arranged themselves by overlap into three distinct features, with wavelengths
of $\sim$5128, 5176, and 5221~\AA. The match of these against features in the
observed spectrum is also indicated in Figure 5, confirming the field strength
determination from the sodium lines.  \ion{Ca}{1} $\lambda$4226 is also present
in absorption, but any magnetic splitting is contained within the broadened
profile. Finally, several sharp features around 8500~\AA\ deserve mention.
While some of these might be interpreted as Zeeman components of \ion{Ca}{2}
$\lambda\lambda8498-8662$, most are artifacts of the data (cf. the spectrum of
SDSS J1036+6522 directly above in Figure 5).  The behavior of the \ion{Ca}{2} H
and K lines has been studied in the appropriate field strength regime (Kemic
1975), but we are not aware of any computations of the infrared triplet.
Unfortunately, LHS 2534 provides no guidance, as the available spectra do not
extend sufficiently far to the red.

We note that the spectrum of SDSS J0157+0033 is strongly blanketed below
5000~\AA\ as indicated by both the spectroscopy and photometric colors.  The
effect is not as dramatic as for LHS 2534, suggesting a somewhat warmer
temperature than that star's $\sim$6000 K.  The lack of \halpha\ in either
star (a dip of dubious significance in SDSS J0157+0033 is centered 23~\AA\ to
the blue of \halpha\ at zero field -- too far to be the quadratic shift at
3.7~MG and too near to be the $\sigma^-$ component) indicates that the
atmospheres are primarily helium, but proper modeling should be carried out to
verify this conclusion.

{\it SDSS J0211+0031 and SDSS J2151+0031:\/} 
Spectral features for each of these stars are weak, owing in part to the rather
low surface temperatures.  However, both are confirmed as magnetic from our
followup circular polarimetry: $v=-2\%\pm0.5\%$ for SDSS J0211+0038 and
$-0.6\%\pm0.2\%$ for SDSS J2151+0031, summed over the range
$\lambda\lambda4200-8200$.  The field strength determinations of 490~MG and
$\sim$300~MG, respectively, are based largely on the presence of a narrow
feature near 5850~\AA\ that we associate with a leisurely turnaround of the
2s0$-$3p0 component of \halpha\ (Figure 2). For SDSS J0211+0038, this
identification is confirmed by several other lines, however the $\lambda$5850
line is extremely weak in SDSS J2151+0031 and our field strength estimate must
be considered tentative pending better quality spectra.

{\it SDSS J0333+0007 and SDSS J0005$-$1002:\/} 
Discovered by Reimers et al. (1998) as HE0330$-$0002 and proven to be magnetic
by Schmidt et al. (2001), this star was recovered in the EDR (G\"ansicke et
al. 2002). The white dwarf probably has a He-dominated atmosphere but the
absorption features, seen in Figure 5, have not been successfully ascribed to
\ion{He}{1}.  Indeed, comparison of the $u-g$ and $g-r$ colors (+0.47 and
+0.13, respectively) to pure helium atmospheric models suggests a temperature
of 7500~K, too cool to produce helium transitions.  Any of a variety of trace
species such as carbon and/or carbon-based molecules are possible.  The
candidate magnetic star SDSS J0005$-$1002 shows two strong troughs between
$4000-4500$~\AA\ as well as weak, diffuse features around 5000~\AA, near the
position of \halpha, and possibly at $\sim$7150~\AA. Several of these
approximate the locations of \ion{C}{1} and \ion{C}{2} lines seen in
relatively hot DQ stars (Liebert 1977; Harris et al. 2003).  The dip around
6580~\AA\ even shows a hint of triplet structure suggesting a mean magnetic
field of $\sim$3~MG.  However, several of the features are weak and lines in
the blue may be affected by flux calibration problems, so further spectroscopy
and polarimetry are required to clarify the nature of this star.

{\it SDSS J0933+0051, SDSS J1113+0146, and SDSS J1333+0016:\/} 
The latter two stars are nearly spectroscopic twins to LP 790-29 and LHS 2229,
respectively (Liebert et al. 1978; Schmidt et al. 1999).  Spectropolarimetry
shown in the same figure demonstrates that, like LP 790-29 and LHS 2229, both
are strongly magnetic, with stronger polarization seen in the bands.  The
extremely deep, overlapping features (Figure 6) are generally reminiscent of
the C$_2$ Swan bands, and Bues (1999) has shown that LP 790-29 can be modeled
using these features in a magnetic field of $\sim$50~MG.  The additional
wavelength shift of $\sim$100~\AA\ to the blue for SDSS J1113+0146 suggests that
its field may be somewhat higher. For LHS 2229 and SDSS J1333+0016, the band
progression is certainly molecular in origin, but the appearance is more
scalloped and the species responsible is not yet determined.  One possibility
is C$_2$H (Schmidt et al. 1999).  SDSS J0933+0051 appears to show the same
structure, but the bands are far weaker.  These stars are also discussed by
Harris et al. (2003).

{\it SDSS J1036+6522:\/} 
The key clues to a magnetic interpretation of SDSS J1036+6522 are the triplets
centered at 4772, 5382, and $\sim$7122~\AA.  In order, these match wavelengths
of \ion{C}{1} transitions: $3s\,^3P^0-4p\,^3P$, $3s\,^1P^0-4p\,^1P$, and a
blend of $3p\,^3D-4d\,^3F^0$ and $3p\,^3D-5s\,^3P^0$.  The splitting of
$\pm$31, 39, and 68~\AA, respectively, implies a mean field strength
$B_s=2.9$~MG, suggesting a polar field strength for an assumed dipolar pattern
of $\sim$4~MG.  Another depression centered near 5050~\AA\ is primarily
\ion{C}{1} $3s\,^1P^0-4p\,^1D$, and additional weaker features are probably
present. SDSS J1036+6522 is the first DQ star that clearly shows
magnetically-split lines of atomic carbon, and thus is an interesting example
of a new magnetic species.  White dwarfs of spectral type DQ -- those with
atomic features as well as the C$_2$ variety -- are generally thought to have
helium-rich atmospheres enriched by convective dredge-up from the carbon core
(e.g., Liebert 1983; Pelletier et al. 1986).

{\it SDSS J2247+1456:\/} 
This star is the highest-field example found thus far in the SDSS and one of
the brightest at $r=17.6$.  In total flux, the widely displaced Balmer
components are reminiscent of those in the most strongly magnetic white dwarf
known, PG 1031+234 (Schmidt et al. 1986).  Followup spectropolarimetry obtained
at the MMT is shown in Figure 7.  Here we see that in circular polarization
SDSS J2247+1456 also closely resembles PG 1031+234, with values reaching nearly
$-10\%$ in the blue and a large amount of structure in the lines.  Faraday
effects are clearly important in the atmosphere, as all features show positive
intrinsic circular polarization regardless of whether they represent $\sigma^+$
components (e.g., most of the lines shortward of 5000~\AA\ and the pair around
7000~\AA) or $\pi$ components (the sharp line due to the turnaround of
$2s0-3p0$ near 5850~\AA; cf. Figure 2).  Such effects are expected (e.g.,
Achilleos \& Wickramasinghe 1995; Wickramasinghe \& Ferrario 2000) and are also
seen in PG 1031+234.  However, adequate modeling of high-field situations has
not yet been carried out, despite the fact that such efforts might well lead to
a better understanding of continuous opacity sources in highly magnetized
atmospheres.

{\it SDSS J2322+0039:\/} 
Our solution for this star, $B_p=19$~MG, $i\sim60^\circ$, is significantly
different from the low-inclination, 13~MG determination of G\"ansicke.  This
is the only such discrepancy for the magnetic DA stars in common between the
analyses.  However, we note that the discrimination between
low-field/low-inclination and higher-field/high-inclination options can be
somewhat problematic with spectra of limited quality, as the two solutions
yield similar mean surface (``effective'') field strengths.

{\it SDSS J2323$-$0046:\/} 
G\"ansicke et al. (2002) interpret the broad depressions near \hbeta\ and
\hgamma\ (Figure 3) as Zeeman-smeared features of hydrogen in a field of
$\sim$30~MG, but point out the existence of narrower lines that may be due to
\ion{He}{1}.  Close inspection reveals that \ion{He}{1} $\lambda$5876 is
actually a Zeeman triplet indicating a field $B_p\sim5$~MG, and $\lambda$6678
is consistent with the same interpretation. Lacking evidence of \halpha, we
interpret the broad features in the blue as confluences of \ion{He}{1} lines,
some of which are indicated in Figure 3. As implied by G\"ansicke et al., it is
conceivable that this star is a DA+DBH binary, and further spectroscopy and
spectropolarimetry should be carried out to evaluate this possibility.

\section{Discussion}

\subsection{Distribution with Temperature}

The lifetime of a low-order magnetic field against free ohmic decay has been
computed to be more than 10$^9$ yr (e.g., Chanmugam \& Gabriel 1972; Muslimov,
Van Horn, \& Wood 1995), and observational support for this result can be found
in the lack of a dependence of field strength on stellar temperature (e.g.,
Wickramasinghe \& Ferrario 2000).  With temperature estimates for the new SDSS
examples covering the range $\sim$7000 K $-$ 30,000 K, this lack of correlation
is emphatically verified, both when the new stars are analyzed alone and when
taken as part of the entire 116-object magnetic sample. It would appear that
the magnetic fields on white dwarfs indeed behave like amplified, fossil
remnants from earlier stages of evolution.

\subsection{Distribution with Field Strength}

The improved statistics provided by the SDSS discoveries warrant a new look at
the distribution of magnetic white dwarfs as a function of field strength.  In
Figure 8 we present this as a histogram {\it vs.\/} polar field strength for
both the SDSS discoveries (solid) and for all 116 examples known at the
present time (hatched).  Bins represent equal intervals in $\log B_p$, and
polar field strengths are taken from spectral models where available.  For
stars that have not yet been successfully interpreted, field strengths are
estimated from the degree of continuum circular polarization.  Such estimates
are particularly crude, but we do not wish to bias the results by omission,
and the number of unsolved stars is sufficiently small that an error of a
factor 2 or so in field strength will not affect the interpretation.

When evaluating the field strength distribution, the relative completeness of
search techniques must be questioned.  Because the discovery of magnetic fields
below $\sim$2~MG requires the rather specialized techniques of
spectropolarimetry, high-resolution spectroscopy of Balmer-line cores, or
asteroseismology, lists are incomplete here.  Thus, we focus attention on field
strengths $B_p>3$~MG, where the SDSS has already doubled the number of known
magnetic stars.  The discovery of white dwarfs of any type by spectroscopic
classification from the SDSS depends in large part on the complicated algorithm
the project uses for selecting the subset of detected sources for
spectroscopy.  An important consideration is that point sources with unusual
colors often receive fibers because one of the SDSS categories is
``serendipity'' (see below).  Highly magnetic white dwarfs with broad,
displaced, and distorted absorption lines may show quite unusual colors (e.g.,
SDSS J2247+1456 or esp. SDSS J1113+0146 and SDSS J1333+0016!) and so be
targeted for spectroscopic followup.  This is particularly true for
$T\lesssim9,000$~K, where the colors of low-to-moderate field magnetic white
dwarfs approach the locus of metal-poor main sequence stars, but magnetic
degenerates with strong absorption features may remain outside of the stellar
locus.  For hotter white dwarfs, the magnetic white dwarfs with fields below
100~MG or so probably have a similar selection probability as nonmagnetic
stars.

Such considerations suggest that the histogram of field strengths plotted in
Figure 8 may actually be biased in favor of the highest fields.  This makes the
apparent peak around 5~MG $-$ 30~MG all the more compelling.  The currently
popular scenario for the origin of magnetic white dwarfs is that they evolve
from magnetic Ap and Bp stars.  Such stars are found to occupy the range 2.8~kG
$<\langle B \rangle<$ 30~kG, and the low-field cutoff is claimed to be {\it
not\/} the result of selection effects (Mathys 2001). If we assume a factor of
10$^4$ for the typical amplification factor due to flux-freezing during
formation of the degenerate core, and that a factor of 2 is reasonable for the
scaling between the mean surface field and the polar field strength of the
equivalent dipole, we find that the Ap/Bp hypothesis is most applicable for the
highest-field magnetic white dwarfs, with $\sim$50~MG $\lesssim B_p \lesssim
500$~MG.  Another origin would seem to be required for the magnetic degenerates
with weaker fields. Because of their comparatively large number, yet-undetected
magnetic F main sequence stars would appear to be attractive candidates.

\subsection{Completeness of the SDSS for Magnetic White Dwarfs}

Some idea of the effectiveness of the fiber assignment and inspection process
for identifying magnetic white dwarfs can be gained by comparing samples of
objects found by different techniques.  First, we have concurred with all of
the magnetic identifications of G\"ansicke et al. (2002), and we have added 2
magnetic stars that were not noted by them but were included in the EDR:
SDSS J1729+5632 is a DA star displaying a moderate field but its spectrum is of
relatively low signal-to-noise ratio, and SDSS J0211+0031 is a high-field DA with
weak, highly-displaced features.

Of greater significance is the recovery rate for known magnetic white dwarfs
whose positions have been included in the survey.  Eight such stars are located
in regions covered by the imaging survey of DR1: KUV03292+0035, HE0330$-$0002,
G111-49, PG1015+015, GD90, G195-19, LBQS1136$-$0132, and Feige 7.
Additionally, LHS~2534 and G99-37 were surveyed but are outside DR1 and
LHS~2534 received a fiber. Six of the previously-known magnetic stars received
fibers and were visually recognized as magnetic by team members who were blind
to the identities of the stars.  Of the remainder, Feige 7, G195-19, and G99-37
exceed the brightness threshold for detector saturation during a typical
spectroscopic exposure and so cannot be fiber candidates (Stoughton 2002).  The
net result -- 5 of the 6 qualifying magnetic white dwarfs within DR1 were
recovered -- suggests that the overall success rate is high for identifying
targets fainter than about magnitude 15 and with temperatures similar to those
that dominate existing white dwarf catalogs.  As might be expected, both
previously-known objects with very odd spectra were recovered: the unexplained
HE0330$-$0002 and the strong-line magnetic DZ LHS~2534 (Figure 5).

The current level of completeness and its prognosis for future releases of the
survey can also be assessed by inquiring into the fiber selection criteria used
for the SDSS magnetic white dwarfs.  Referring to the distribution among SDSS
target selection categories (e.g., Stoughton et al. 2002), 22 of the 60 stars
fell into the QSO region of color space, and an additional 13 stars were chosen
as HOT\usp STD (hot subdwarf standard).  Thus, nearly 60\% of the magnetic
white dwarfs appeared in high-priority categories for which spectroscopic
fibers are nearly assured.  The remaining targets fell off the stellar locus in
categories SERENDIPITY\usp BLUE (14 stars), SERENDIPITY\usp DISTANT (7), or
were correctly picked out as hot white dwarfs in STAR\usp WHITE\usp DWARF (4
stars).  These three classes receive rather low priority for spectroscopy, and
the resulting sky coverage is highly nonuniform.  However, if we assume that a)
60\% of magnetic white dwarfs will always appear in the spectroscopic database,
b) the remaining 40\% have possibly a 50:50 chance of being targeted, and c)
that manual inspection is very efficient at recognizing magnetic spectra, we
can broadly account for the high fraction of recovered magnetic stars.  Of
course, future results will be subject to changes in the procedures, including
any refinement of the color criteria for the various object categories.

Finally, photometry and astrometric information is available for the imaged,
but unrecovered, magnetic stars.  For completeness this information is included
as Table~5. Note that G195-19 in particular is so bright that the $gri$
magnitudes are uncertain.

\section{Summary and Conclusions}

Despite the fact that the SDSS is tailored to extragalactic criteria, the
survey is proving to be a rich source of unusual stellar objects.  With roughly
one quarter of the eventual sky coverage now completed, the known list of
magnetic white dwarfs has nearly doubled in size, and new discoveries include
the first low-field DB stars as well as magnetic examples of exotic atmospheric
compositions like atomic and molecular carbon and metallic-line white dwarfs.
Due to our very limited knowledge of the magnetic behavior of atomic and
molecular species, a few high-field objects continue to frustrate efforts for
line identification.

The list of magnetic+nonmagnetic white dwarf binaries that was previously
comprised of LB11146 (Liebert et al. 1993), G62-46 (Bergeron, Ruiz, \& Leggett
1993), and RX J0317$-$855 (=EUV0317$-$855; e.g., Barstow et al. 1995; Vennes et
al. 2003) has grown by two.  Both SDSS J0042+0019 and SDSS J0847+4842 deserve
followup study to determine their periods and component masses.  No
double-magnetic systems have yet been found, and no detached magnetic white
dwarf+main-sequence systems have been identified.  The evolutionary
implications of these results are not yet clear, but the rapid spin rate, high
temperature, and extraordinarily large mass of RX J0317$-$855 suggest that
magnetic fields may be acquired during the course of stellar evolution, at
least when mergers are involved.

The efficiency of the overall identification pipeline for magnetic white dwarfs
is high for reasonably hot stars with $B \gtrsim 3$~MG and $g \gtrsim 15$.
The improved statistics and more uniform selection process provided by the
SDSS has yielded a distribution of magnetic white dwarfs that peaks in the
range $\sim$5~MG $-$ 30~MG and shows a tail to several hundred MG.  While it
is reasonable to assume that the highest-field magnetic white dwarfs evolve
from main-sequence Ap/Bp stars, the bulk of magnetic degenerates in this peak
and below would appear to require an alternate source.  Yet undetected
magnetic F stars, whose convective envelopes would destroy an ordered
underlying field structure with $B\sim1$~G $-$ 3~kG, would seem to be likely
candidates.

\acknowledgements{Funding for the Sloan Digital Sky Survey has been provided by
the Alfred P. Sloan Foundation, the Participating Institutions, the National
Aeronautics and Space Administration, the National Science Foundation, the U.S.
Department of Energy, the Japanese Monbukagakusho, and the Max Planck Society.
The SDSS is a joint project of The University of Chicago, Fermilab, the
Institute for Advanced Study, the Japan Participation Group, The Johns Hopkins
University, Los Alamos National Laboratory, the Max-Planck-Institute for
Astronomy (MPIA), the Max-Planck-Institute for Astrophysics (MPA), New Mexico
State University, University of Pittsburgh, Princeton University, the United
States Naval Observatory, and the University of Washington.  We are grateful to
the Directors of the Steward and MMT Observatories for awarding the observing
time to follow up these discoveries, to M. Strauss for noting some of the
magnetic candidates, and to P. Smith for assistance with the observations.
Special thanks go to D.T. Wickramasinghe for general discussions about highly
magnetic stars.  Support was provided by NSF grant 97-30792.}

\clearpage

\begin{deluxetable}{crrrrrl}
\tablenum{1}
\tablewidth{6.3in}
\setlength{\tabcolsep}{0.06in}
\tablecaption{SDSS DA MAGNETIC WHITE DWARFS}
\tablehead{\colhead{Star}
& \colhead{Pl.-MJD-Fib.}
& \colhead{UT Date}
& \colhead{UT}
& \colhead{$B_p$}
& \colhead{$i$}
& \colhead{Comment}
\\
\colhead{(SDSS+)}
& & &
& \colhead{(MG)}
& \colhead{($^\circ$)}}
\startdata
J004248.19+001955.3 &   690-52261-594  &   2001-12-18  &    3:12  &       14  &      30 & DAH+DA pair; not in DR1 \\
J021116.34+003128.5 &   405-51816-382  &   2000-09-29  &    9:58  &      490  & \nodata & \\
J030407.40$-$002541.7 & 411-51817-172  &   2000-09-30  &   10:37  &       11  &      60 & (1: 10.8 MG, 50$^\circ$) \\
J033145.69+004517.0 &   415-51810-370  &   2000-09-23  &    8:34  &       12  &      60 & KUV03292+0035; (1: 12.1 MG, 55$^\circ$) \\
J034308.18$-$064127.3 & 462-51909-117  &   2000-12-31  &    5:16  &       13  &      45 & \\
J034511.11+003444.3 &   416-51811-590  &   2000-09-24  &    9:35  &      1.5  &       0 & (1: 1.5 MG, 0$^\circ$) \\
J075819.57+354443.7 &   757-52238-144  &   2001-11-25  &    8:12  &       27  &      30 & \\
J075959.56+433521.3 &   437-51869-369  &   2000-11-21  &   10:18  &      220  & \nodata & G111-49 \\
J080743.33+393829.2 &   545-52202-009  &   2001-10-20  &   11:00  &       49  &      30 & \\
J084155.74+022350.6 &   564-52224-248  &   2001-11-08  &   11:09  &        6  &      90 & \\
J084716.21+484220.4 &   550-51959-629  &   2001-02-19  &    4:40  &  $\sim$3  & \nodata & DAH+DB pair \\
J085830.85+412635.1 &   830-52293-070  &   2002-01-19  &    5:44  &        2  &      30 & not in DR1 \\
J092527.47+011328.7 &   475-51965-315  &   2001-02-25  &    7:53  &      2.2  & \nodata & \\
J100005.67+015859.2 &   500-51994-557  &   2001-03-26  &    2:47  &       20  &      30 & \\
J101618.37+040920.6 &   574-52355-166  &   2002-03-21  &    4:09  &      7.5  &      30 & not in DR1 \\
J101805.04+011123.5 &   503-51999-244  &   2001-03-26  &    4:19  &      120  & \nodata & PG1015+015 \\
J105404.38+593333.3 &   561-52295-008  &   2002-01-21  &    9:05  &       17  &      90 & not in DR1 \\
J105628.49+652313.5 &   490-51929-205  &   2001-01-15  &    9:17  &       28  &      60 & \\
J111010.50+600141.4 &   950-52378-568  &   2002-04-14  &    7:56  &      6.5  &      70 & not in DR1 \\
J112852.88$-$010540.8 & 326-52375-565  &   2002-04-11  &    4:41  &        3  &      60 & not in DR1 \\
J113357.66+515204.8 &   879-52365-586  &   2002-04-01  &    7:11  &      7.5  &      90 & not in DR1 \\
J113839.51$-$014903.0 & 327-52294-583  &   2002-01-12  &   10:29  &       24  &      60 & LBQS1136$-$0132; not in DR1 \\
J114006.37+611008.2 &   776-52319-042  &   2002-02-14  &    9:29  &       58  &      20 & not in DR1 \\
J115418.14+011711.4 &   515-52051-126  &   2001-05-20  &    4:22  &       32  &      45 & \\
J115917.39+613914.3 &   777-52320-069  &   2002-02-15  &    9:41  &     15.5  &      60 & not in DR1 \\
J121209.31+013627.7 &   518-52282-285  &   2002-01-08  &   12:01  &       13  &      80 & not in DR1 \\
J121635.37$-$002656.2 & 288-52000-276  &   2001-04-01  &    6:24  &       61  &      90 & (1: 63 MG, 75$^\circ$) \\
J122209.44+001534.0 &   289-51990-349  &   2001-03-22  &    6:25  &       14  &      80 & (1: 12 MG, 40$^\circ$) \\
J124851.31$-$022924.7 & 337-51997-264  &   2001-03-29  &    7:36  &        7  &      40 & \\
J133340.34+640627.4 &   603-52056-112  &   2001-05-27  &    5:20  &       13  &      60 & \\
J134043.10+654349.2 &   497-51989-182  &   2001-03-21  &    9:14  &        3  &      60 & \\
J144614.00+590216.7 &   608-52081-140  &   2001-06-21  &    4:03  &        7  &      70 & \\
J151745.19+610543.6 &   613-52345-446  &   2002-03-09  &   12:19  &       17  &      30 & not in DR1 \\
J153532.25+421305.6 &  1052-52466-252  &   2002-07-10  &    5:04  &      4.5  &      60 & not in DR1 \\
J153829.29+530604.6 &   795-52378-637  &   2002-04-14  &   10:27  &       12  &      30 & not in DR1 \\
J154213.48+034800.4 &   594-52045-400  &   2001-05-16  &    8:47  &        8  &      60 & \\
J160437.36+490809.2 &   622-52054-330  &   2001-05-25  &    8:48  &       53  &       0 & \\
J165203.68+352815.8 &   820-52438-299  &   2002-06-13  &    8:14  &      9.5  &      60 & not in DR1 \\
J172045.37+561214.9 &   367-51997-461  &   2001-03-22  &   11:51  &       21  &      30 & (1: 21 MG, 85$^\circ$) \\
J172329.14+540755.8 &   359-51821-415  &   2000-10-03  &    4:00  &       35  &      10 & (1: 33 MG, 35$^\circ$) \\
J172932.48+563204.1 &   358-51818-239  &   2000-10-01  &    2:16  &       28  & \nodata & \\
J204626.15$-$071037.0 & 635-52145-227  &   2001-08-24  &    4:40  &        2  &      60 & \\
J205233.52$-$001610.7 & 982-52466-019  &   2002-07-11  &    7:08  &       13  &      80 & not in DR1 \\
J214930.74$-$072812.0 & 644-52173-350  &   2001-09-21  &    4:37  &       42  &      30 & \\
J215135.00+003140.5 &   371-52078-500  &   2001-06-12  &    9:37  &  $\sim$300 & \nodata & \\
J215148.31+125525.5 &   733-52207-522  &   2001-10-25  &    4:21  &       21  &      90 & not in DR1 \\
J221828.59$-$000012.2 & 374-51791-583  &   2000-09-04  &    5:19  &      225  &      30 & \\
J224741.46+145638.8 &   740-52263-444  &   2001-12-20  &    2:50  &      560  & \nodata & not in DR1 \\
J232248.22+003900.9 &   383-51818-421  &   2000-10-01  &    4:29  &       19  &      60 & (1: 13 MG, 25$^\circ$) \\
\enddata
\tablerefs{
(1) G\"ansicke et al. (2002)}
\end{deluxetable}

\clearpage

\begin{deluxetable}{crrrrrl}
\tablenum{2}
\tablewidth{5.2in}
\setlength{\tabcolsep}{0.06in}
\tablecaption{SDSS DB MAGNETIC WHITE DWARFS}
\tablehead{\colhead{Star}
& \colhead{Pl.-MJD-Fib.}
& \colhead{UT Date}
& \colhead{UT}
& \colhead{$B_p$}
& \colhead{$i$}
& \colhead{Comment}
\\
\colhead{(SDSS+)}
& & &
& \colhead{(MG)}
& \colhead{($^\circ$)}}
\startdata
J001742.44+004137.4  & 687-52518-510  &   2002-09-01   &   9:44   &     8.3  &      90 & not in DR1 \\
J014245.37+131546.4  & 429-51820-311  &   2000-10-01   &   6:18   &       4  &      60 & \\
J232337.55$-$004628.2 & 383-51818-215 &   2000-10-01   &   4:29   &     4.8  &      30 & (1) \\
\enddata
\tablerefs{
(1) G\"ansicke et al. (2002)}
\end{deluxetable}

\begin{deluxetable}{crrrrl}
\tablenum{3}
\tablewidth{6.2in}
\setlength{\tabcolsep}{0.06in}
\tablecaption{ADDITIONAL SDSS MAGNETIC WHITE DWARFS}
\tablehead{\colhead{Star}
& \colhead{Pl.-MJD-Fib.}
& \colhead{UT Date}
& \colhead{UT}
& \colhead{$B_s$}
& \colhead{Comment}
\\
\colhead{(SDSS+)}
& & &
& \colhead{(MG)} }
\startdata
J000555.91$-$100213.4 & 650-52143-037  &   2001-08-22  &   7:37  & ? & DQ? \\
J015748.15+003315.1 &   700-52199-627  &   2001-10-16  &   8:30  & 3.7 &  DZ; not in DR1 \\
J033320.37+000720.7 &   415-51810-492  &   2000-09-23  &   8:34  & ? &  HE0330$-$0002; (1) \\
J093313.14+005135.4 &   475-51965-003  &   2001-02-25  &   7:53  & ? & resembles LHS2229 \\
J103655.38+652252.0 &   489-51930-520  &   2001-01-21  &   9:17  & 4: &  DQ; (2) \\
J111341.33+014641.7 &   510-52381-184  &   2002-04-17  &   4:20  & ? &  resembles LP790-29; not in DR1 \\
J121456.39$-$023402.8 & 333-52313-399  &   2002-02-07  &  11:54  & 1.9 & DZ; LHS 2534; not in DR1 \\
J133359.86+001654.8 &   298-51662-484  &   2000-04-28  &   5:47  & ? &  resembles LHS2229 \\
\enddata
\tablerefs{
(1) G\"ansicke et al. (2002); (2) Liebert et al. (2003)}
\end{deluxetable}

\clearpage

\begin{deluxetable}{lrrrrrr}
\tablenum{4}
\tablewidth{4.55in}
\tablecaption{PHOTOMETRY OF SDSS MAGNETIC WHITE DWARFS}
\tablehead{\colhead{Star (SDSS+)}
& \colhead{$u$}
& \colhead{$g$}
& \colhead{$r$}
& \colhead{$i$}
& \colhead{$z$}
& \colhead{$T_{\rm eff}$ (K)}}
\startdata
J000555.91$-$100213.4 &    17.31 & 17.69 & 18.11 & 18.46 & 18.76 & \nodata \\
J001742.44+004137.4 &      16.86 & 16.96 & 17.21 & 17.44 & 17.71 & 15,000 \\
J004248.19+001955.3 &      19.79 & 19.48 & 19.58 & 19.68 & 19.73 & 11,000 \\
J014245.37+131546.4 &      17.61 & 17.69 & 17.99 & 18.20 & 18.44 & 15,000 \\
J015748.15+003315.1 &      21.22 & 19.53 & 19.20 & 19.23 & 19.32 & \nodata \\
J021116.34+003128.5 &      18.66 & 18.55 & 18.52 & 18.63 & 18.73 & 9,000:\\
J030407.40$-$002541.7 &    18.06 & 17.80 & 17.93 & 18.11 & 18.35 & 11,500 \\
J033145.69+004517.0 &      17.31 & 17.22 & 17.47 & 17.72 & 18.00 & 15,500 \\
J033320.37+000720.7 &      17.03 & 16.56 & 16.43 & 16.42 & 16.52 & 7,500: \\
J034308.18$-$064127.3 &    19.49 & 19.47 & 19.60 & 19.81 & 19.92 & 13,000: \\
J034511.11+003444.3 &      19.18 & 18.66 & 18.53 & 18.50 & 18.52 & 7,500 \\
J075819.57+354443.7 &      18.07 & 18.18 & 18.56 & 18.83 & 19.01 & 22,000 \\
J075959.56+433521.3 &      16.76 & 16.19 & 16.23 & 16.21 & 16.09 & 9,000: \\
J080743.33+393829.2 &      20.41 & 20.14 & 20.33 & 20.57 & 20.60 & 13,000 \\
J084155.74+022350.6 &      19.57 & 18.99 & 18.80 & 18.72 & 18.77 & 7,000 \\
J084716.21+484220.4 &      17.86 & 17.88 & 18.20 & 18.48 & 18.68 & 19,000 \\
J085830.85+412635.1 &      17.73 & 17.05 & 16.89 & 16.83 & 16.88 & 7,000: \\
J092527.47+011328.7 &      19.04 & 18.59 & 18.69 & 18.76 & 18.86 & 10,000 \\
J093313.14+005135.4 &      19.80 & 19.59 & 19.28 & 19.33 & 19.45 & \nodata \\
J100005.67+015859.2 &      20.39 & 20.03 & 20.04 & 20.06 & 20.28 & 9,000 \\
J101618.37+040920.6 &      20.56 & 20.30 & 20.36 & 20.49 & 20.33 & 10,000 \\
J101805.04+011123.5 &      16.52 & 16.28 & 16.44 & 16.55 & 16.74 & 12,500: \\
J103655.38+652252.0 &      18.31 & 18.52 & 18.85 & 19.15 & 19.26 & \nodata \\
J105404.38+593333.3 &      20.58 & 20.23 & 20.28 & 20.45 & 20.40 & 9,500 \\
J105628.49+652313.5 &      19.83 & 19.70 & 19.99 & 20.31 & 20.49 & 16,500 \\
J111010.50+600141.4 &      17.68 & 17.96 & 18.43 & 18.73 & 19.04 & 30,000 \\
J111341.33+014641.7 &      18.62 & 19.21 & 18.47 & 18.26 & 18.10 & \nodata \\
J112852.88$-$010540.8 &    20.66 & 20.41 & 20.51 & 20.77 & 20.61 & 11,000 \\
J113357.66+515204.8 &      17.26 & 17.34 & 17.71 & 18.01 & 18.33 & 22,000 \\
J113839.51$-$014903.0 &    17.97 & 17.62 & 17.73 & 17.95 & 18.18 & 10,500 \\
J114006.37+611008.2 &      20.02 & 19.64 & 19.90 & 20.06 & 20.20 & 13,500 \\
J115418.14+011711.4 &      17.46 & 17.75 & 18.15 & 18.45 & 18.80 & 27,000: \\
J115917.39+613914.3 &      18.87 & 18.96 & 19.39 & 19.66 & 19.85 & 23,000 \\
J121209.31+013627.7 &      18.43 & 17.99 & 18.07 & 18.24 & 18.40 & 10,000 \\
J121456.39$-$023402.8 &    20.90 & 18.31 & 17.74 & 17.55 & 17.50 & \nodata \\
J121635.37$-$002656.2 &    19.86 & 19.60 & 19.87 & 20.09 & 20.11 & 15,000 \\
J122209.44+001534.0 &      20.56 & 20.27 & 20.50 & 20.67 & 21.21 & 14,000 \\
J124851.31$-$022924.7 &    18.71 & 18.38 & 18.62 & 18.87 & 19.19 & 13,500 \\
J133340.34+640627.4 &      18.16 & 17.88 & 18.10 & 18.25 & 18.56 & 13,500 \\
J133359.86+001654.8 &      19.06 & 19.41 & 18.33 & 18.06 & 18.16 & \nodata \\
J134043.10+654349.2 &      18.74 & 18.42 & 18.74 & 18.95 & 19.23 & 15,000 \\
J144614.00+590216.7 &      20.55 & 20.10 & 20.31 & 20.44 & 20.59 & 12,500 \\
J151745.19+610543.6 &      20.85 & 20.50 & 20.55 & 20.74 & 21.17 & 9,500 \\
J153532.25+421305.6 &      20.46 & 20.36 & 20.74 & 20.89 & 22.04 & 18,500 \\
J153829.29+530604.6 &      19.53 & 19.29 & 19.49 & 19.67 & 20.03 & 13,500 \\
J154213.48+034800.4 &      19.56 & 19.11 & 19.11 & 19.22 & 19.43 & 8,500 \\
J160437.36+490809.2 &      18.22 & 17.90 & 17.91 & 18.00 & 18.15 & 9,000 \\
J165203.68+352815.8 &      19.68 & 19.25 & 19.42 & 19.54 & 19.75 & 11,500 \\
J172045.37+561214.9 &      19.99 & 20.10 & 20.47 & 20.72 & 21.30 & 22,000 \\
J172329.14+540755.8 &      19.10 & 18.78 & 18.85 & 19.01 & 19.27 & 10,000 \\
J172932.48+563204.1 &      20.24 & 20.02 & 20.08 & 20.21 & 20.87 & 10,500 \\
J204626.15$-$071037.0 &    18.32 & 17.99 & 17.90 & 17.95 & 18.02 & 8,000 \\
J205233.52$-$001610.7 &    18.44 & 18.50 & 18.80 & 19.09 & 19.44 & 19,000 \\
J214930.74$-$072812.0 &    17.36 & 17.43 & 17.80 & 18.12 & 18.36 & 22,000: \\
J215135.00+003140.5 &      18.16 & 17.84 & 17.84 & 17.87 & 18.03 & 9,000: \\
J215148.31+125525.5 &      18.30 & 18.11 & 18.32 & 18.58 & 18.80 & 14,000 \\
J221828.59$-$000012.2 &    18.29 & 18.06 & 18.35 & 18.54 & 18.71 & 15,500: \\
J224741.46+145638.8 &      17.31 & 17.38 & 17.62 & 17.95 & 18.20 & 18,000: \\
J232248.22+003900.9 &      18.96 & 19.22 & 19.43 & 19.69 & 19.85 & 20,000: \\
J232337.55$-$004628.2 &    17.88 & 18.02 & 18.31 & 18.54 & 18.77 & 15,000 \\
\enddata
\end{deluxetable}

\begin{deluxetable}{lcrrrrr}
\tablenum{5}
\tablewidth{5.1in}
\tablecaption{UNRECOVERED MAGNETIC WHITE DWARFS IN IMAGED AREAS\tablenotemark{a}}
\tablehead{\colhead{Star}
& \colhead{SDSS Coordinates (J2000)}
& \colhead{$u$}
& \colhead{$g$}
& \colhead{$r$}
& \colhead{$i$}
& \colhead{$z$} \\
& \colhead{($hh$:$mm$:$ss.ss$ $\pm dd$:$mm$:$ss.s$)}}
\startdata
Feige 7         & 00:43:45.98 $-$10:00:25.1 & 14.25 & 14.40 & 14.75 & 15.02 & 15.32 \\
G99-37          & 05:51:19.49 $-$00:10:20.6 & 15.37 & 14.72 & 14.39 & 14.32 & 14.38 \\
GD90            & 08:19:46.35 $+$37:31:27.7 & 16.10 & 15.73 & 15.85 & 16.02 & 16.28 \\
G195-19         & 09:15:55.97 $+$53:25:23.0 & 14.49 & 13.99: & 14.08: & 13.84: & 13.98 \\
\tablenotetext{a}{All but GD90 exceed the brightness threshold for being assigned a spectroscopic fiber.}
\enddata
\end{deluxetable}

\clearpage

\begin{figure}
\plotone{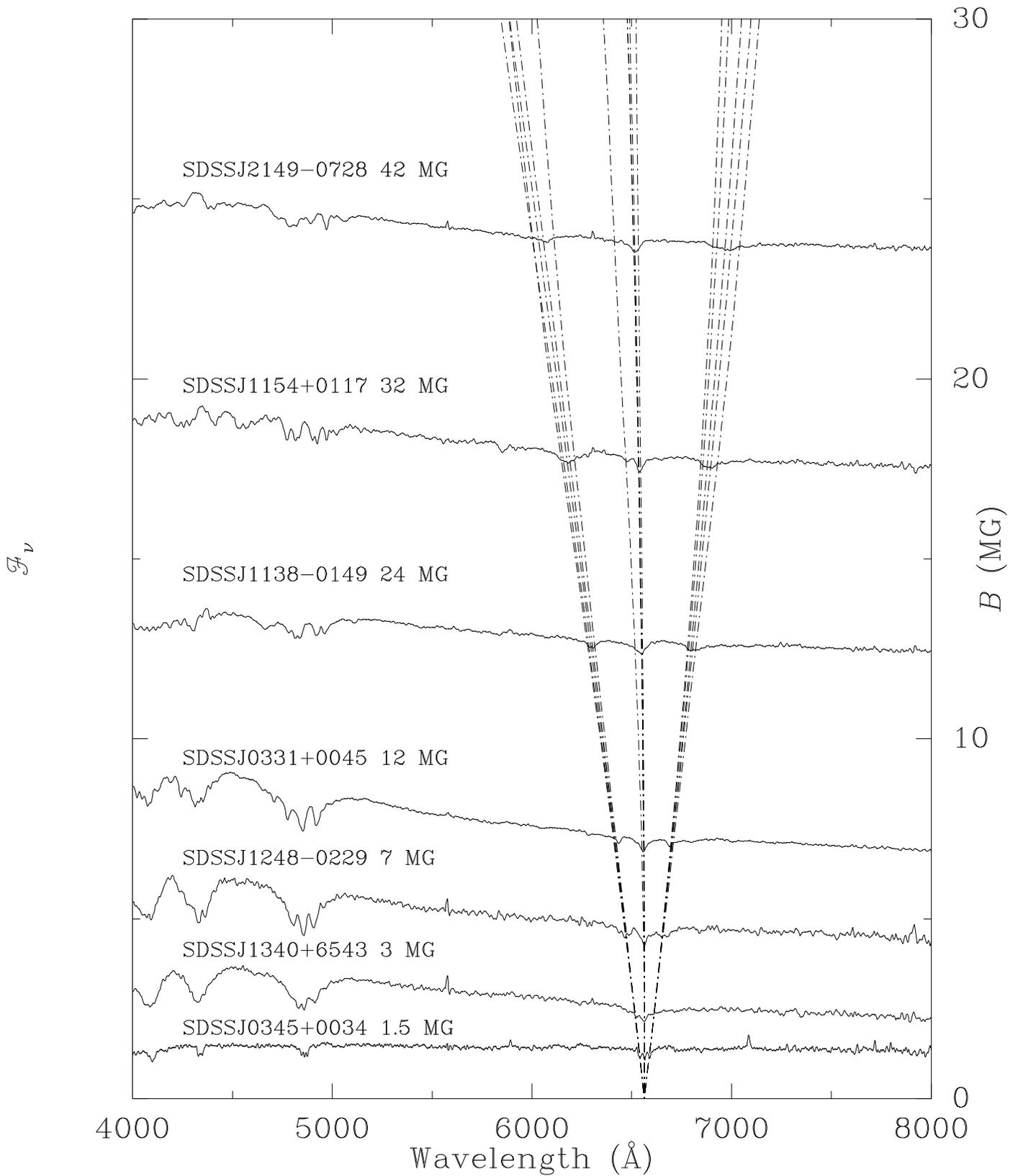}

\figcaption{A sample of the DR1 magnetic DA white dwarfs showing polar fields
from 1.5~MG to 42~MG as indicated.  For brevity, stars are indicated by
SDSS J$hhmm{\pm}ddmm$. Spectra are positioned along the ordinate according to
approximate mean surface field strength, $B_s$, in order to match absorption
features with the locations of Zeeman components of \halpha\ for $0<B<30$~MG.}

\end{figure}

\clearpage

\begin{figure}
\plotone{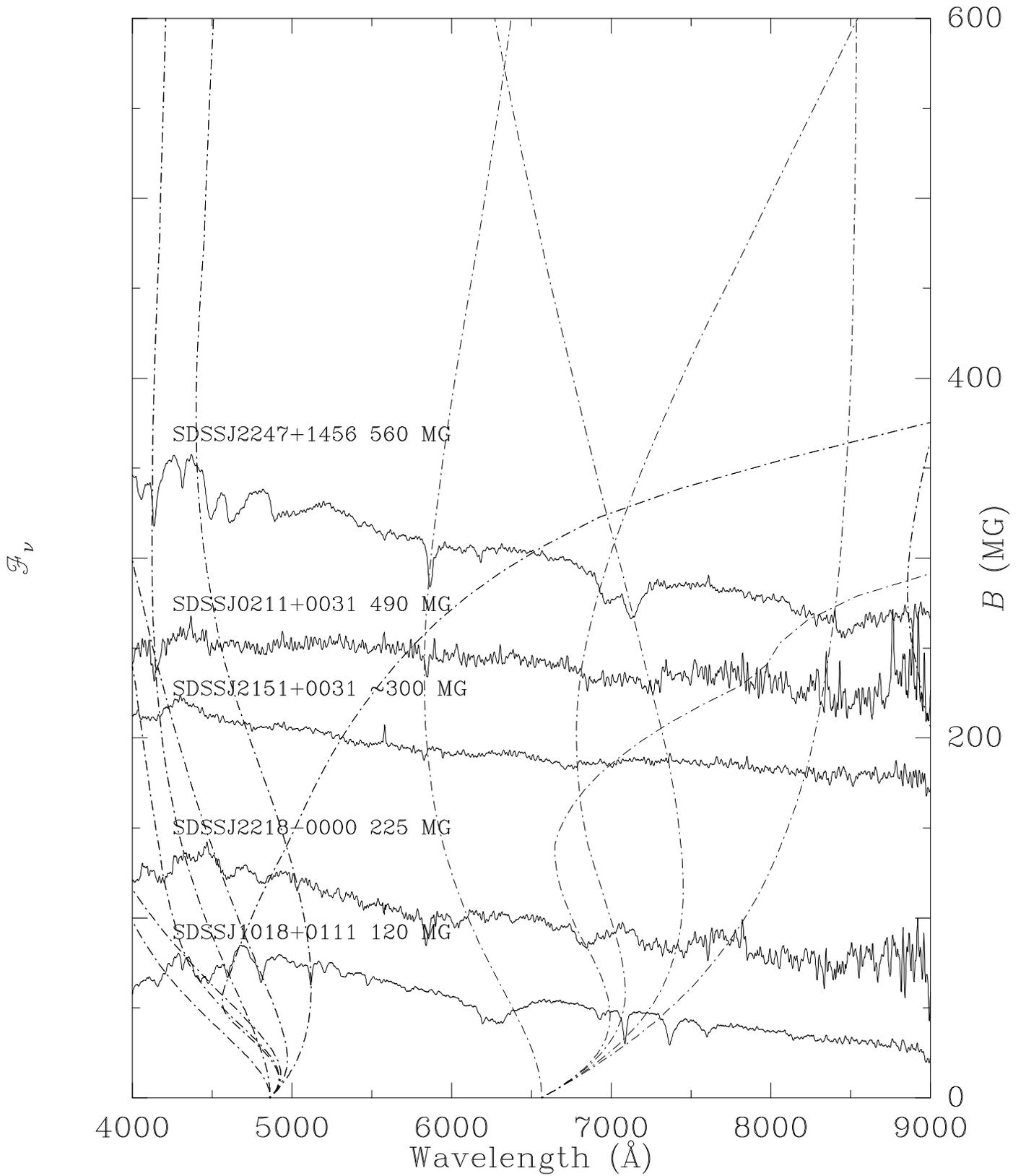}

\figcaption{Magnetic DA white dwarfs from the DR1 with polar field strengths
$B_p=100$~MG $-$ 500~MG compared to those components of the Balmer series that
undergo turnarounds or reach stationary points in the region of interest.
Spectra are positioned along the ordinate by the approximate mean surface field
strength $B_s$.}

\end{figure}

\clearpage

\begin{figure}
\plotone{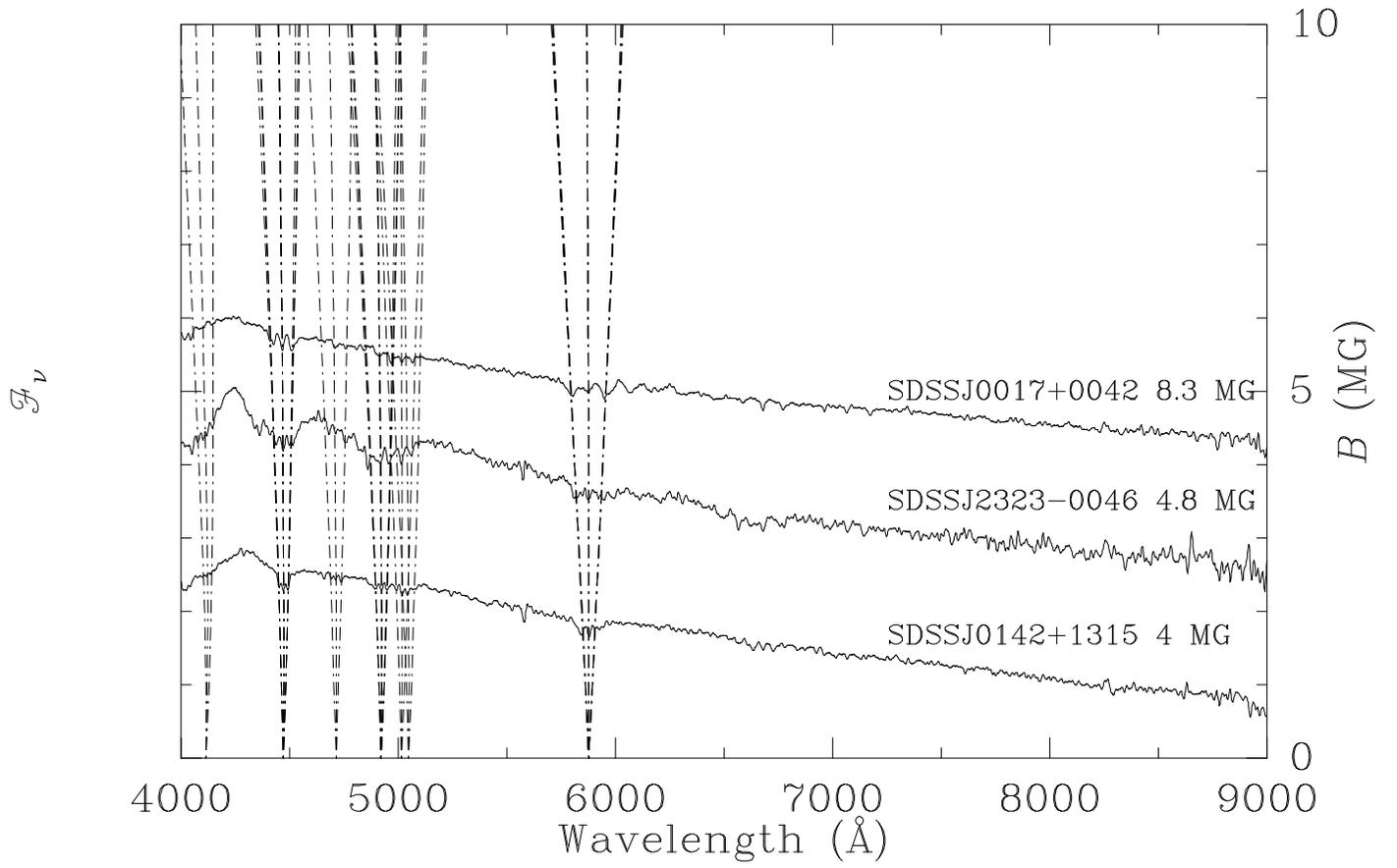}

\figcaption{As in Figures 1 and 2 for magnetic DB white dwarfs with
$B_p<10$~MG.  The behavior of several \ion{He}{1} lines is also shown for
comparison.}

\end{figure}

\clearpage

\begin{figure}
\plotone{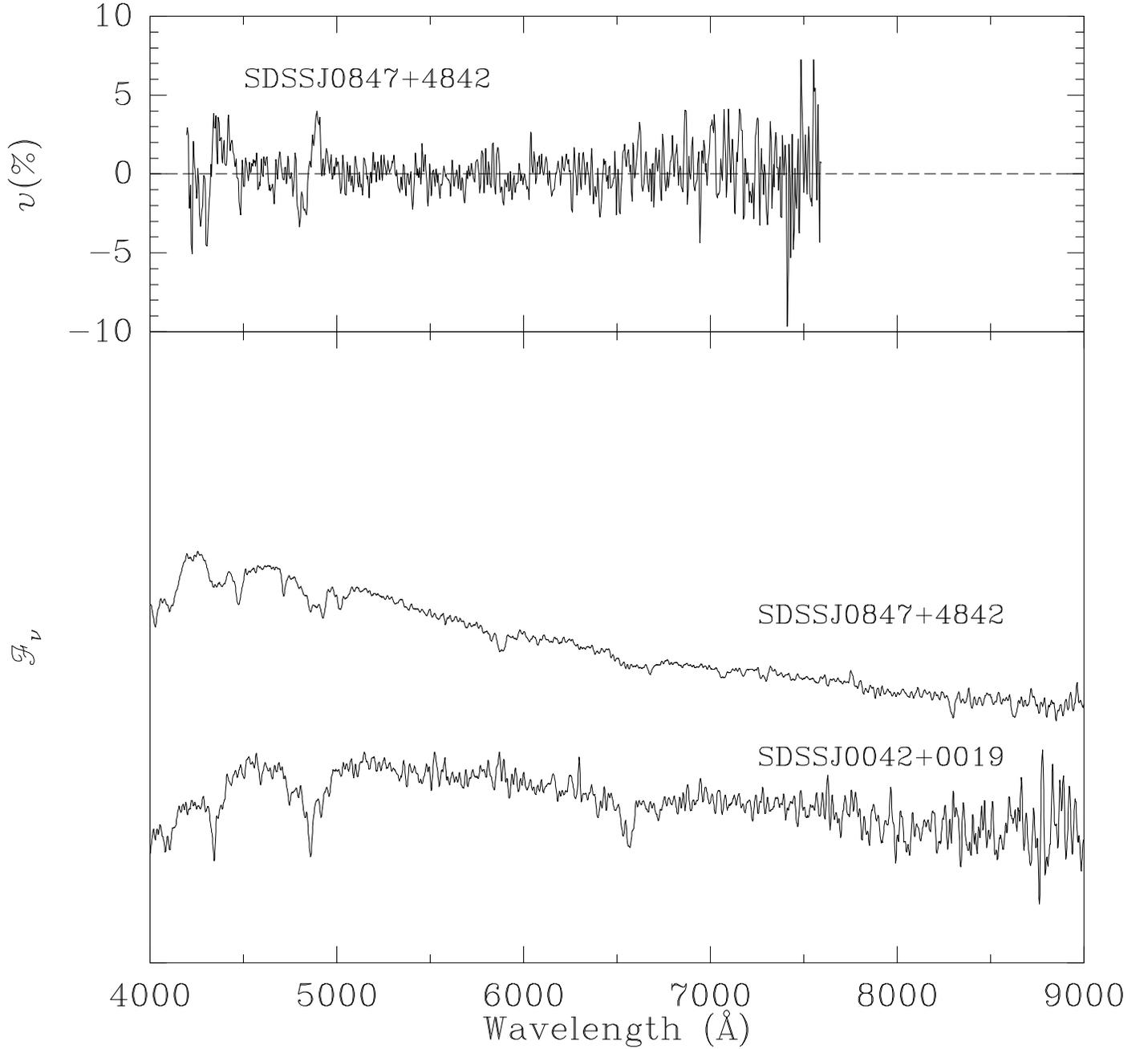}

\figcaption{New magnetic + nonmagnetic white dwarf binaries. {\it (Bottom):\/}
The DAH+DA pair SDSS J0042+0019 with $B_p=14$~MG and the apparent DAH+DB
SDSS J0847+4842 with $B_p\sim3$~MG.  Note the breadth of, e.g., \halpha\ vs. the
\ion{He}{1} lines in the latter star. {\it (Top):\/} Circular
spectropolarimetry of SDSS J0847+4842 showing polarization reversals across
\hbeta\ and \hgamma, but not across the \ion{He}{1} lines.}

\end{figure}

\clearpage

\begin{figure}
\plotone{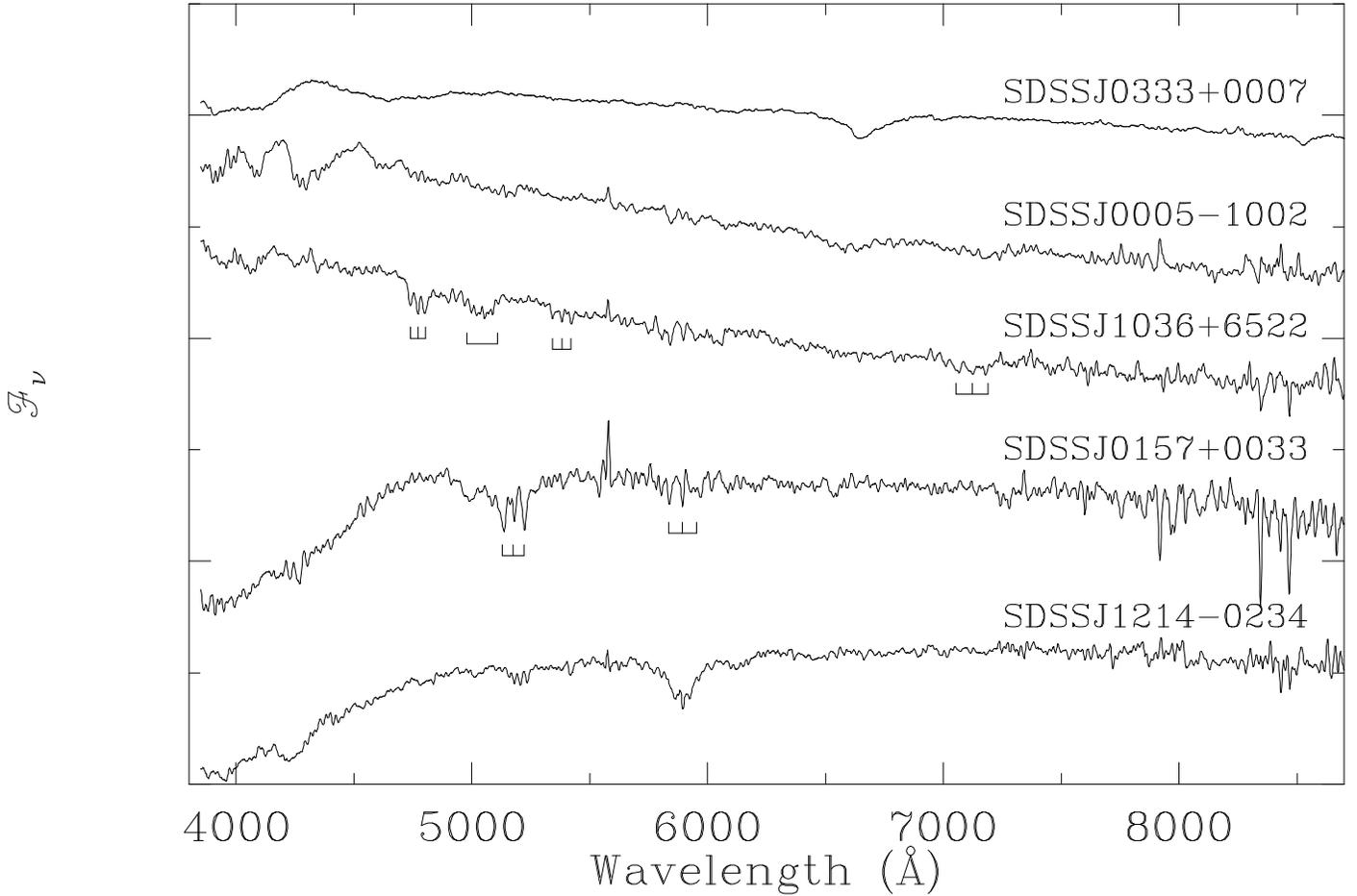}

\figcaption{SDSS spectra of magnetic white dwarfs and white dwarf candidates of
various types.  SDSS J1214$-$0234 (=LHS 2534) and SDSS J0157+0033 are magnetic DZ
stars showing Zeeman-split lines of \ion{Na}{1} and \ion{Mg}{1} at $B_s=1.9$
and 3.7~MG, respectively.  SDSS J1036+6522 is the first magnetic DQ star to show
split \ion{C}{1} lines.  SDSS J0005$-$1002 may be similar in nature, but requires
confirmation.  Spectropolarimetry of SDSS J0333+0007 (=HE0330$-$0002) proves it
to be magnetic, but the distorted features do not correspond with a known
atomic or molecular species. Data artifacts affect the spectra of the faintest
stars for $\lambda\gtrsim8000$~\AA.}

\end{figure}

\clearpage

\begin{figure}
\plotone{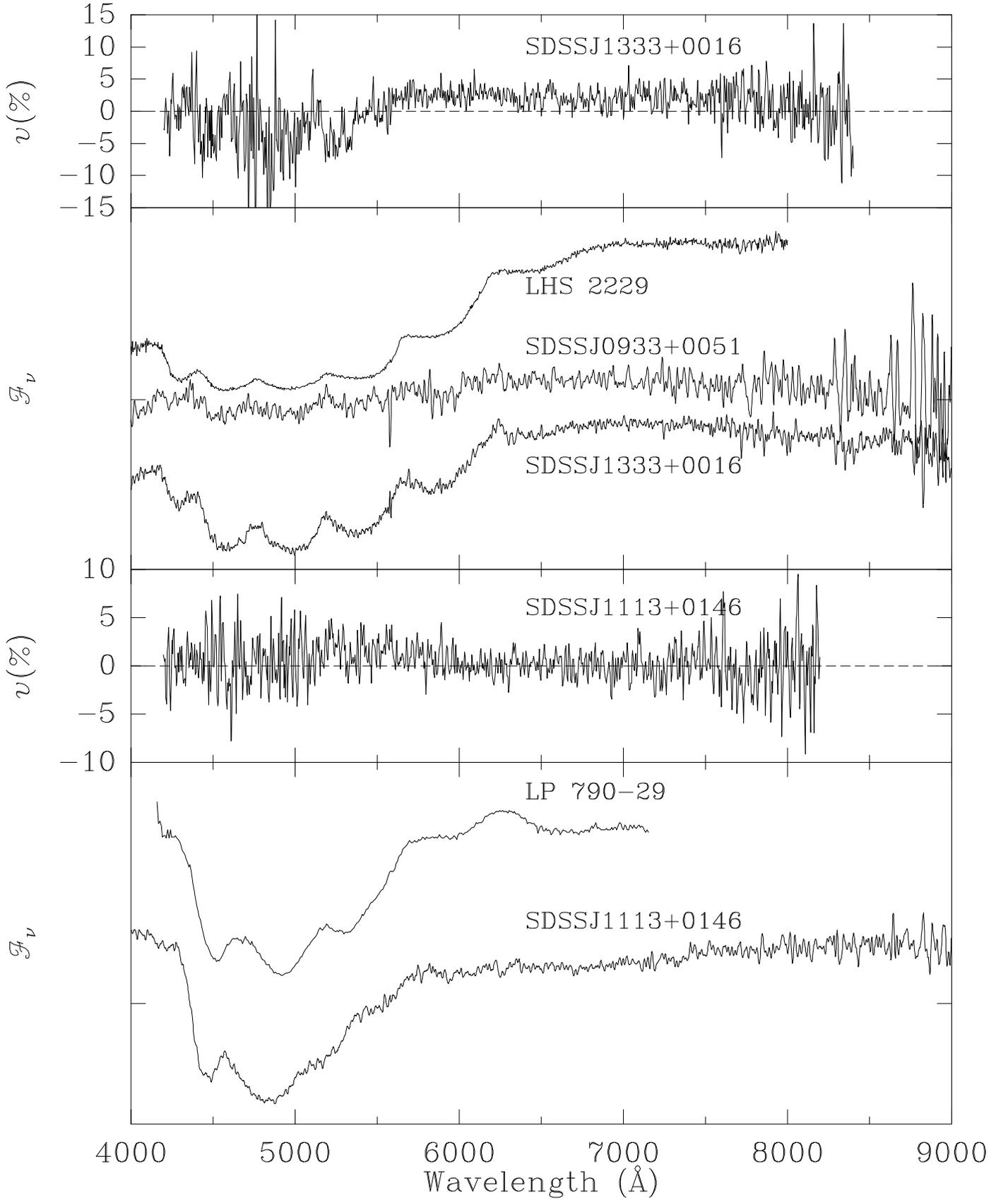}

\figcaption{New SDSS white dwarfs showing molecular features compared to
previously-known examples with very similar spectra.  The spectra of LHS 2229
and LP 790-29 have been displaced upward for clarity. Spectropolarimetry
demonstrates that SDSS J1113+146 and SDSS J1333+0016 are magnetic, with the
features in SDSS J1113+0146 and LP 790-29 likely being the C$_2$ Swan bands.
The species responsible for the remarkable series in SDSS J1333+0016 and LHS
2229 are not yet known.  SDSS J0933+0051 appears to exhibit the same band
structure, but the features are much weaker.}

\end{figure}

\clearpage

\begin{figure}
\plotone{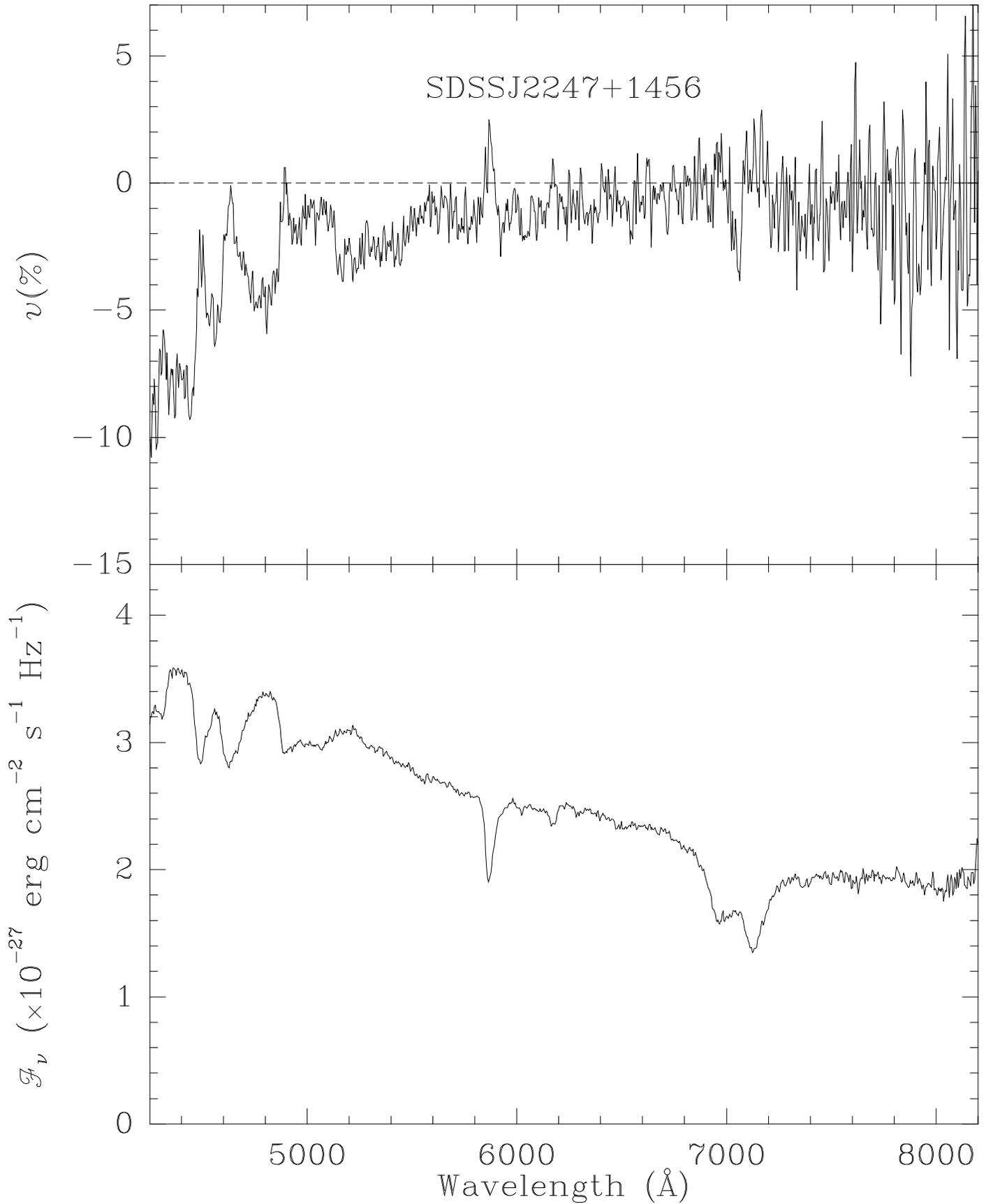}

\figcaption{Circular polarization {\it (top)\/} and total flux {\it (bottom)\/}
spectra of a new hydrogen-atmosphere SDSS magnetic white dwarf with
$B_p\sim560$~MG.}

\end{figure}

\clearpage

\begin{figure}
\includegraphics{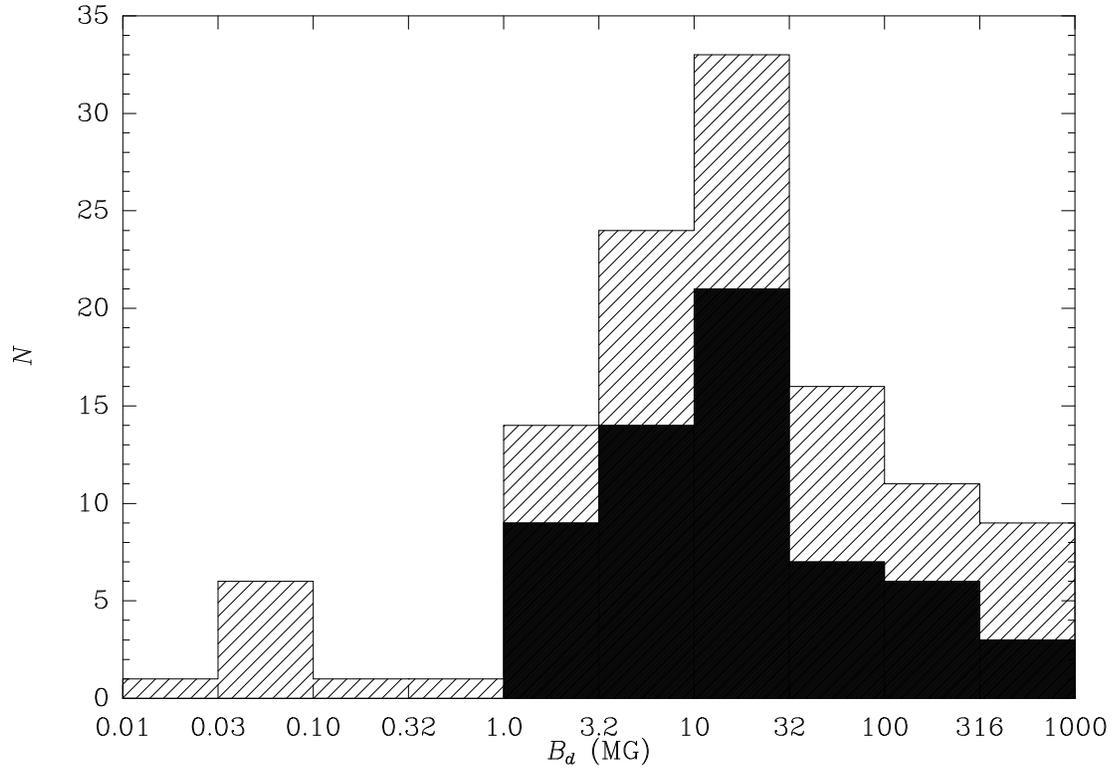}
\vspace{3.truein}

\figcaption{Histogram of magnetic white dwarfs in equal intervals of $\log B$.
SDSS discoveries are shaded black; the distribution for all known magnetic
white dwarfs is hatched.  Selection effects are not likely to affect the
overall distribution of stars with $B_p\gtrsim3$~MG, so the general peak in the
range $\sim$$5-30$~MG is considered to be real.}

\end{figure}

\end{document}